# High-performance THz metamaterial absorber

Jianfei Zhu,[1] Zhaofeng Ma,[1] Wujiong Sun,[2] Fei Ding,[1] Qiong He,[2] Lei Zhou,[2] and Yungui Ma[1,*]

[1]*State Key Laboratory for Modern Optical Instrumentation, Centre for Optical and Electromagnetic Research, Department of Optical Engineering, Zhejiang University, Hangzhou 310058, China*

[2]*State Key Laboratory of Surface Physics and Key Laboratory of Micro and Nano Photonic Structures (Ministry of Education), and Physics Department, Fudan University, Shanghai 200433, China*

We demonstrated an ultra-broadband, polarization-insensitive and wide-angle metamaterial absorber for terahertz (THz) frequencies using arrays of truncated pyramid unit structure made of metal-dielectric multilayer composite. In our design each sub-layer behaving as an effective waveguide is gradually modified in their lateral width to realize a wideband response by effectively stitching together the resonance bands of different waveguide modes. Experimentally, our five layer sample with a total thickness 21μm is capable of producing a large absorptivity above 80% from 0.7 to 2.3 THz up to the maximum measurement angle 40°. The full absorption width at half maximum (FWHM) of our device is around 127%, greater than those previously reported for THz frequencies. Our absorber design has high practical feasibility and can be easily integrated with the semiconductor technology to make high efficient THz-oriented devices.

*Corresponding author: yungui@zju.edu.cn

Electromagnetic metamaterials (MMs) have attracted great research attentions in the last decades for their exotic dynamic properties that are absent in nature [1,2]. Many electromagnetic (EM) effects and devices deemed impossible before have been proposed and demonstrated employing the unique characteristics of MMs; as typically represented by negative refraction [3-5], perfect imaging lens [6] and invisible cloak [7,8]. For most of their applications, material loss and bandwidth will be the very critical parameters to evaluate the performance of these devices made of MMs, whose resonance nature often seriously limits the real applications of these devices. But on the other side, material loss will be a gain parameter for wave absorption generally relied by EM stealth [9] and many probe technologies [10]. In 2008 Landy et al. first proposed a thin perfect metamaterial absorber (MA) simultaneously exciting electric and magnetic resonances to realize the impedance match with the surrounding air, thus eliminating any boundary reflection [11]. After this, different sets of MA designs with matched impedance have been investigated in a wide frequency region ranging from microwave to optics [12-19]. However, these perfect absorbers usually work at limited frequencies due to their resonance features and may not suit many practical applications that have bandwidth requirements. Efforts have been made to broaden the absorption spectrum, e.g., by modeling a gradually changed cone-shaped topology to slow down the incident waves and totally absorb them, as recently reported in microwave [20] and optics using metallic film stacks [21].

In this work we implement a metal-dielectric multilayer composite to realize an ultra broadband THz wave absorber that can work with weak polarization and incident angle dependence. Terahertz is a spectrum that has many important technological applications such as imaging [22], security detection [23], or wireless communication [24], which all requires a perfect absorber to efficiently collect wave energy. Artificial THz metamaterial absorber was firstly designed in 2008 by Tao et al. using a two-layer metal resonator that was implemented at 1.3 THz with 70% absorptivity [12], and later the same authors enhanced the angle performance of their THz MA by an ELC metamaterial array [25]. In 2010 Ye et al. proposed a three-layered metal-cross structure to acquire a relatively large band of THz absorption with a 40% of theoretical FWHM [14]. Similar stacking designs have also been proposed using other types of metallic elements to improve the bandwidth of a THz absorber [26-28], among which a large 48% of FWHM has been experimentally obtained [27]. In our model, wideband absorption is obtained by stitching the resonance absorption of a series of stacked parallel-plate metallic waveguides whose varying effective lengths provide multiple resonance sub-bands in a composite structure. During the experiment we have overcome the manufactured difficulty often encountered for a three dimensional THz device and has demonstrated a 21-um-thick MA with absorptivity above 80% and FWHM of 127 % in the range of 0.7 to 2.3 THz frequencies up to our maximum wave incident angle 40°. Our methodology for THz absorber may be coupled to find important applications in some THz-oriented devices such as a microbolometer element sensor for THz imaging [29].

Figure 1 gives the schematic of our THz MA device and the unit cell structure. We adopt a square lattice for our sample in order to acquire dual polarization working performance. The lateral lattice constant is chosen to be 95 μm, which is smaller than the largest wavelength discussed here. Each unit cell consists of five layers of 200-nm-thick square metal films stacked along the vertical direction. All the metal films are embedded in a dielectric insulating matrix at a fixed distance 4 μm. We gradually tune the metal patch width of each layer from the bottom to the top to form a truncated pyramid structure with an optimized constant difference 4 μm (the width of the bottom patch is 70 μm). Our design has taken into account the practical feasibility for the THz sample although a pure pyramid structure with more shrunk metal patches may better satisfy the impedance-match condition [20,21]. In this work photoresist SU-8 with a measured permittivity $\varepsilon = 2.79 - i\, 0.31$ as the insulating matrix and chromium (conductivity $\sigma = 8 \times 10^6$ s/m) as the metallic layer. In the bottom a 200 nm thick opaque chromium film is pre-deposited as the background mirror.

Numerical simulation of our MA sample is first performed within CST Microwave Studio. Periodic boundary conditions are used for a unit cell in the *x* and *y* directions, as shown in Fig. 1, and a plane wave in the *x-z* plane which is assumed to illuminate the sample from the top at an angle $\theta$ with the normal (or *z*-axis) direction. In simulation we examine the incident situations with both transverse electric (TE) and transverse magnetic (TM) polarizations. Figures 2(a)-2(d) show the simulated absorptivity of our MA model as a function of frequency and incident angle. At normal incidence, high absorptivity larger than 80% is achieved in a wide frequency range of 0.9-1.6 THz. The FWHM is calculated to be 72%, which explicitly implies a very good wideband property. For TE polarization, as shown in Figs. 2(a) and 2(b), the absorption intensity drops when the incident angle increases larger than 30° but the bandwidth seems to experience little changes. For TM polarization, as shown in Figs. 2(c) and 2(d), the absorption intensity has weak dependence on the incident angle even up to 70°. But for TM polarization, we see two narrow absorption branches at the frequencies above 1.6 THz, which might be attributed to higher order Bloch modes excited by our structures with a period comparable with the applied wavelengths at higher frequencies. The relatively broad and flat absorption peak accords with the typical characteristics of a Fabry-Perot-like metallic waveguide mode whose resonance frequency is primarily dependent on the waveguide dimension [30].

Next we resort to EM field patterns to have a deep understanding on the underlying mechanism of our broadband MA. Firstly, we recall that a single layer array of metal patch on a conducting background spaced by a thin dielectric will effectively behave as horizontal waveguides and strongly interact with the incident wave at the excitation of waveguide harmonic modes. It will lead to strong wave attenuation at the resonances of these modes by transferring EM energy into heat [31]. Here we conduct an experiment to confirm this point. Figure 3 plots the measured and simulated absorption lines for a single layer of metal patch array with period 140 μm, patch width 105 μm, and SU-8 thickness 4 μm. A narrow band absorption peak around 0.8 THz is obtained at the resonance of the first order parallel-plate waveguide mode as shown by its electric field pattern given in the inset of Fig. 3. Obviously this absorption peak can be tuned by modifying the length of waveguide or patch width. Here our motivation is to acquire a wideband absorption by favorably stitching together the individual resonance of each parallel-plate waveguide stacked along the vertical direction. It can be realized if any anti-phase coupling between neighboring waveguides, which reduces interaction with the incident wave, can be elegantly suppressed [30].

Figures 4(a)-4(h) show the electric and magnetic field distributions at different frequencies at normal incidence for our wideband MA taken on the central cross-section of a unit cell in the *x-z* plane. It can be seen that the first order parallel-plate waveguide mode is excited in different layers at different frequency values and the mode with a lower frequency is excited in a shorter waveguide. The complementary spatial positions for the electric and the magnetic fields in accord with the standing wave characteristic of a waveguide mode and have little changes at moderate oblique incidence [30]. In principle a large number of stacked waveguides may have the potential to incorporate many resonance modes and realize wave absorption with frequency band as wide as possible. Here our five-layer design is a first technical attempt to demonstrate this technological potential in THz waves.

Our design is implemented by a standard optical lithography process by applying the same structural and material parameters used in the above simulation, i.e., [200-nm-thick chromium/4-μm-thick SU-8 polymer]$_5$. Special carefulness has been taken to fabricate this complex multilayer system, for the metal-polymer composite is very easy to crack and peel off from the substrate in the repetition of depositing and patterning films. Experimentally a 2 cm × 2 cm sample is fabricated (see the inset of Fig. 5) and measured by a commercial THz time domain spectroscopy (Zomega-Z3 system) with a frequency range from 0.2 to 3 THz in TM mode. In the measurement an opaque gold film is used as a reference reflection mirror. Figure 5 gives the measured absorption lines at four wave incident angles: 0, 20, 30 and 40 degrees, for TM polarization. We see that large absorptivity above 80% is achieved in a wide frequency range from 0.7 to 2.3 THz, which is remained up to our maximum measurement angle 40 degrees. At normal incidence the FWHM is measured to be 127%, which is much larger than the experimental value 48% reported in a recent work also for THz waves [27]. It is noted that the measured absorption bandwidth is two times broader than the simulated, as shown in Fig. 2(d), which may be addressed by two reasons: (i) the real material parameters (metal conductivity or dielectric permittivity) may have certain differences with those used in simulation, which is highly possible for our complex multilayer sample where roughness and stress may seriously influence the film qualities, and (ii) the slightly misaligned metal stacks (see the inset of Fig. 5) may introduce additional wave modes to interact with the incident wave. Further deeper investigation on structural and material distributions would be needed to totally understand these influential parameters and their weights. Nonetheless, our measured results unambiguously prove the absorption performance of the truncated pyramid metamaterial for THz wave application.

In conclusion, we have demonstrated an ultra-broadband THz metamaterial absorber by integrating multiple parallel-plate waveguides in one unit cell and merging together their resonance bands. The gradually modified square patch design also provides a wide incident wave angle and negligible polarization-sensitivity. Novel device applications such as a microbolometer element sensor for THz imaging can be envisioned by coupling the designs of our monolith absorber arrays [29].

**Acknowledgement:** The authors are grateful to the partial supports from NSFCs 61271085, 60990322, 60990321, 11174055 and 11204040, the National High Technology Research and Development Program (863 Program) of China (No. 2012AA030402), the Program of Zhejiang Leading Team of Science and Technology Innovation, NCET, MOE SRFDP of China, the Fundamental Research Funds for the Central Universities, the Program of Shanghai Subject Chief Scientist

**Captions**

FIG. 1 (Color online) Schematic of our THz MA design. Left: a top view of the structure and right: an oblique view of unit cell. Our design includes a square array (period 95 μm in the x-y plane) of truncated pyramid element made of five metal films stacked along the z-axis and in between separated by a 4-μm-thick dielectric polymer. In our prototype the square metallic patch from the top to the bottom has a side width of 54, 58, 62, 66 and 70 μm. The sample has a total thickness of 21 μm. A 200-nm-thick opaque metal wall is used as the background mirror. The incident wave is assumed in the x-z plane at an incident angle, θ.

FIG. 2. (Color online) Simulated absorptivity. (a) and (c) are for TE polarization; (b) and (d) are for TM polarization. In (a) and (c) the x and y axes represent frequency and incident angle, respectively, and the absorptivity value is represented by different colors. (b) and (d) plot the absorptivity lines at selected incident angles of 0, 30, 50 and 70 degrees.

FIG. 3. (Color online) Absorption of one layer of metal patch array. The red (blue) line represents the simulation (measured) result. The inset gives the electric field pattern at the resonance frequency 0.8 THz.

FIG. 4. (Color online) Simulated electric (left panel) and magnetic field (right panel) distributions of our wideband MA at different frequencies. The field patterns are obtained on the central cross section of a unit cell. The top to the bottom row corresponds to the working frequency of 0.99, 1.21, 1.32 and 1.48 THz, respectively. The normalized field magnitude is presented by different colors which have the same definition with those used in Figs. 2(a) and 2(c).

FIG. 5. (Color online) Measured absorptivity of our broadband THz MA at four incident angles: 0, 20, 30 and 40 degrees. The inset gives a top view of our sample whose parameters have been given in Fig. 1 and its caption.

Figure 1

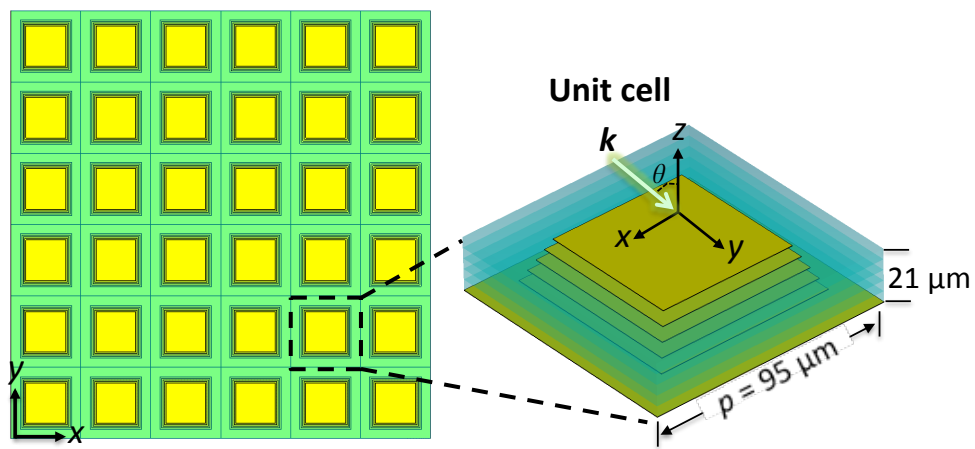

Figure 2

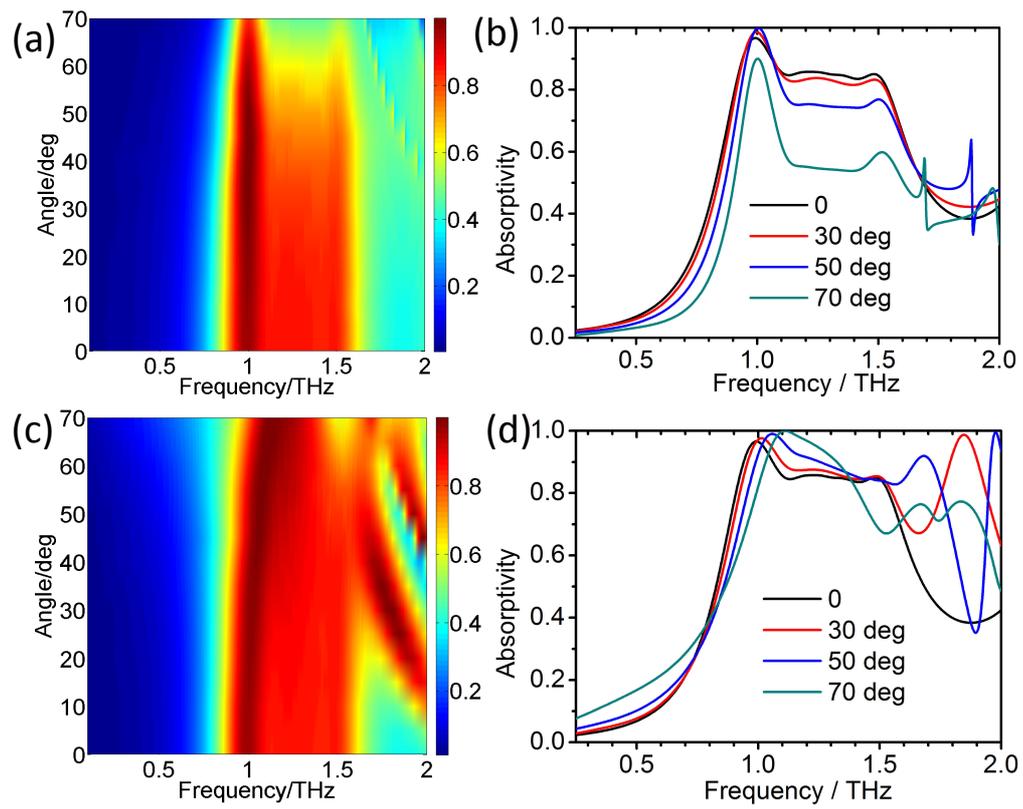

Figure 3

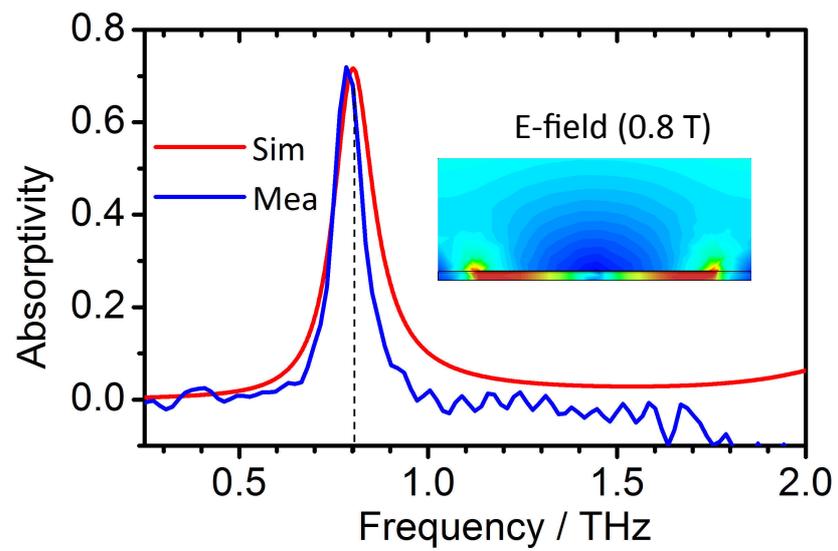

Figure 4

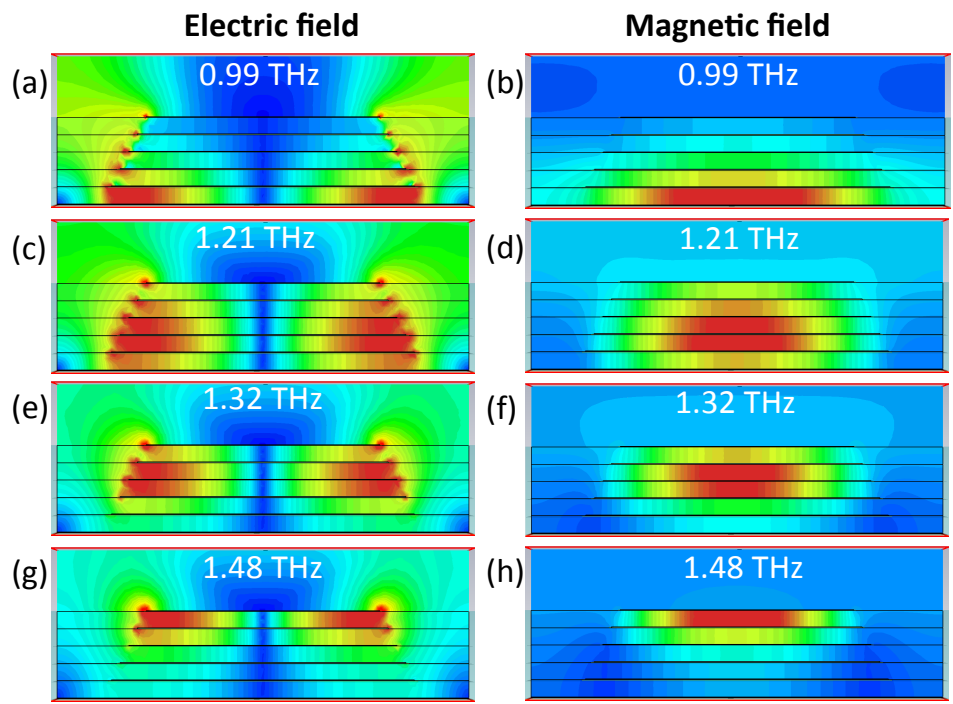

Figure 5

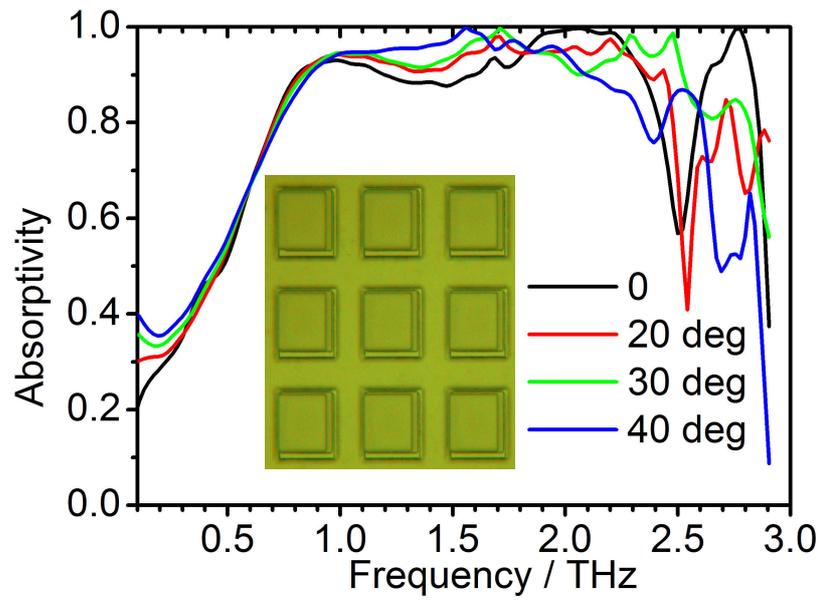